\documentclass[aps,notitlepage]{revtex4-1}
\usepackage{amsmath}
\usepackage{epsfig}
\usepackage{multirow}
\def\bea#1\eea{\begin{align}#1\end{align}}

\newcommand{\bef}{\begin{figure}[h!tb]\centering}
\newcommand{\eef}{\end{figure}}

\begin{document}
\title{Initial-state cold nuclear matter energy loss effects on inclusive jet production 
\\ in p+A collisions at RHIC and LHC}

\date{\today}

\author{Zhong-Bo Kang}
\email{zkang@lanl.gov}
\affiliation{Theoretical Division,
                   Los Alamos National Laboratory,
                   Los Alamos, NM 87545, USA}
                   
\author{Ivan Vitev}
\email{ivitev@lanl.gov}
\affiliation{Theoretical Division,
                   Los Alamos National Laboratory,
                   Los Alamos, NM 87545, USA}                  

\author{Hongxi Xing}
\email{hxing@lanl.gov}
\affiliation{Theoretical Division,
                   Los Alamos National Laboratory,
                   Los Alamos, NM 87545, USA}

\begin{abstract}
Recent measurements of the centrality and rapidity dependence of single inclusive jet production in p+Pb collisions at the LHC have revealed large and non-trivial nuclear modification of the production cross section for this process. In this paper, we explore to what extent such nuclear modification can be understood by the framework of standard cold nuclear matter effects, in particular initial-state cold nuclear matter energy loss. We demonstrate quantitatively that  theoretical calculations which include medium-induced radiative corrections can describe rather reasonably the attenuation of the jet production yields in the large transverse momentum region in d+Au collisions at RHIC and p+Pb collisions at the LHC for central to semi-central collisions. We further show that the observed scaling behavior of the nuclear modification factor as a function of the total jet energy $p_T\, {\cosh} (y)$  for various rapidity intervals has a natural explanation in the picture of cold nuclear matter energy loss. On the other hand, the observed enhancement in peripheral collisions is not described in this picture and could have a different origin. 
\end{abstract}

\maketitle

\section{Introduction}
High energy jet production in p+A collisions has  long  been regarded as a valuable baseline for studies of jet quenching and for constraining the properties of the hot dense medium created in A+A collisions at RHIC and LHC~\cite{Abreu:2007kv,Albacete:2013ei}, since final-state effects associated with the quark-gluon plasma (QGP) are expected to be absent/suppressed. However, recent experimental results on the centrality dependence of high energy jet cross sections in p+Pb collisions at LHC~\cite{Adam:2015hoa,ATLAS:2014cpa} and in d+Au collisions at RHIC~\cite{Perepelitsa:2013jua,Adare:2015gla} seem to challenge the expectation of negligible or small cross section modification. Highly nontrivial and large nuclear effects from different centrality selections are observed at all transverse momenta $p_T$ at forward (in the direction of the proton beam) rapidities and for large $p_T$ at mid-rapidity, and they are manifest as  suppression of the jet yield in central events and enhancement in peripheral collisions~\cite{ATLAS:2014cpa}. 

So far there is no theoretical approach which can successfully describe all  aspects of these surprising results. In particular, they are challenging to explain within the framework of nuclear parton distribution functions (nPDFs)~\cite{Eskola:2009uj}, which follows the usual leading-twist perturbative QCD factorization~\cite{Collins:1989gx} and attributes all nuclear effects to universal nPDFs in the large nucleus. There are also other attempts to understand the experimental data. For example, it was hypothesized~\cite{Bzdak:2014rca} that the observed effects might arise from a suppression of the soft particle multiplicity in events with high energy jets, though the detailed mechanism of such suppression needs further detailed investigation. The correlation between the hard process rates and the soft event activity has also been studied in~\cite{Perepelitsa:2014yta}. Independently, it has also been argued~\cite{Alvioli:2013vk,Alvioli:2014sba,Alvioli:2014eda} that parton configurations in the projectile proton containing a parton with large momentum fraction $x$ interact with a nuclear target with a reduced cross section, which may give rise to the observed effects. At the same time, studies of kinematic bias on the centrality selection of jet events in p+A collisions at the LHC have been performed in~\cite{Armesto:2015kwa}. Although the model is not based on any dynamical mechanism and fails in the case of peripheral collisions, it does capture the trend in the data at the LHC from central to semi-central collisions. 

In this paper, our goal is to explore to what extent the nuclear modification of jet production in p+A collisions observed at both RHIC and LHC can be described by the standard cold nuclear matter (CNM) effects.  In p+A collisions, the production yield of single inclusive jets is affected by a number of CNM effects, all of which have clear physical/dynamical origin, mostly centered around the idea of multiple parton scattering. They include dynamical nuclear shadowing~\cite{Qiu:2004qk,Qiu:2004da}, Cronin effect~\cite{Gyulassy:2002yv,Accardi:2002ik}, incoherent multiple scattering~\cite{Kang:2013ufa,Kang:2014hha} and CNM energy loss~\cite{Gavin:1991qk,Vogt:1999dw,Johnson:2000ph,Kopeliovich:2005ym,Vitev:2007ve,Neufeld:2010dz,Wang:2001ifa,Xing:2011fb,Arleo:2012hn}, which can all be calculated from the elastic, inelastic and coherent scattering processes of partons in large nuclei~\cite{Vitev:2003xu,Vitev:2006bi,Kang:2012kc,Sharma:2012dy,Andronic:2015wma}. It was found in our previous studies that dynamical shadowing, incoherent multiple scattering as well as Cronin effects play important roles only in the small and intermediate transverse momentum regions~\cite{Kang:2012kc}. At very large transverse momentum $p_T$, only CNM energy loss effects that follow the energy dependence  obtained in~\cite{Vitev:2007ve,Neufeld:2010dz}  are relevant in modifying particle production yields 
 in p+A collisions. Therefore,  results on high $p_T$ jets in p+A collisions should provide further important insights into the nature of CNM energy loss and differentiate between theoretical models.

To this end, we study the phenomenological consequence of CNM energy loss effects on the high $p_T$ jet production in p+A reactions. We compare our theoretical calculations to single inclusive jet production in d+Au collisions at RHIC at $200$ GeV~\cite{Perepelitsa:2014yta} and in p+Pb collisions at LHC at $5.02$ TeV~\cite{ATLAS:2014cpa} from central to forward rapidity. We find that CNM energy loss leads to significant suppression of jet production in p+A collisions which is qualitatively consistent with both RHIC and LHC experimental data. In particular, the observed scaling behavior of the nuclear modification factor as a function of the total jet energy for various rapidity intervals finds its 
origin in the CNM energy loss picture. However, the observed enhancement in peripheral collisions is not described in this picture and needs to be further explored. 

The rest of our paper is organized as follows. In Sec.~II, we first review the perturbative QCD formalism for single inclusive jet production in p+p collisions. We then discuss how the cold nuclear matter energy loss is implemented in this formalism. In Sec.~III, we present our phenomenological results of high $p_T$ jet production in d+Au collisions at RHIC and p+Pb collisions at the LHC. We demonstrate quantitatively to what extent the cold nuclear matter energy loss effects can explain the experimental data. We summarize our paper in Sec.~IV.

\section{Cold nuclear matter energy loss in p+A collisions}

To leading order (LO) in the framework of factorized perturbative QCD, single inclusive jet production in p+p collisions, $p(P_a)+p(P_b)\to jet(y, p_T)+X$, can be written as~\cite{Owens:1986mp},
\bea
\frac{d\sigma^{pp}_{\rm jet}}{dyd^2p_T}=\frac{\alpha_s^2}{s}\sum_{a,b}\int\frac{dx_a}{x_a}f_{a/p}(x_a, \mu)\int\frac{dx_b}{x_b}f_{b/p}(x_b, \mu)
H_{ab\rightarrow c}(\hat s,\hat t, \hat u)\delta(\hat s+\hat t+\hat u),
\label{eq:p+p}
\eea
where $y$ and $p_T$ are the rapidity and transverse momentum of the jet, respectively, $\sum_{a,b}$ represents the sum over all parton flavors, $s=(P_a+P_b)^2$ is the center-of-mass energy squared and $\hat s$, $\hat t$, $\hat u$ are standard Mandelstam variables defined at the partonic level. $f_{a/p}(x, \mu)$ are the parton distribution functions with $\mu$ being the factorization scale. $H_{ab\rightarrow c}$ are the short-distance hard-part functions for two partons of flavor $a$ and $b$ to reproduce a jet (at LO it is just the leading parton $c$), which are listed in Ref.~\cite{Kang:2011bp}. In principle, a LO calculation of single inclusive jet production has certain limitations since only at next-to-leading order (NLO) the QCD structure of the jet starts to play a role in the theoretical description of physical observables~\cite{Ellis:1990ek,deFlorian:2013qia}, for example a LO calculation cannot be used to study the dependence of CNM effects on the jet cone radius $R$. Unlike final-state quark-gluon plasma effects, however, the 
dependence of initial-state effects on the jet cone radius is expected to be very small~\cite{He:2011sg}. 
Furthermore, quantitatively the NLO and LO cross sections are very similar for intermediate jet radii $R\sim 0.4$. For this reason LO calculations are very useful to discuss qualitatively the cold nuclear matter effects~\cite{Vitev:2009rd}. At the same time in the large $p_T$ region, it has been shown that the dependence of nuclear modification on the jet cone size is significantly reduced compared to that in the small and moderate $p_T$ region. In this paper, we focus on the study of jets with very large transverse momentum $p_T$, and we thus take the advantage of the LO calculations for simplicity.

\begin{figure}[!t]
\begin{center}
\psfig{file = 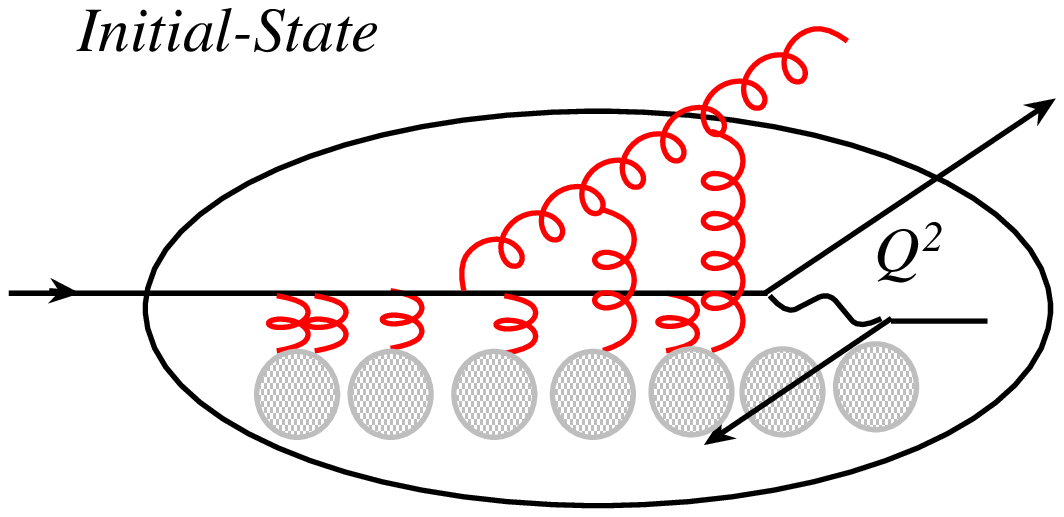,width=2.8in}
\psfig{file = 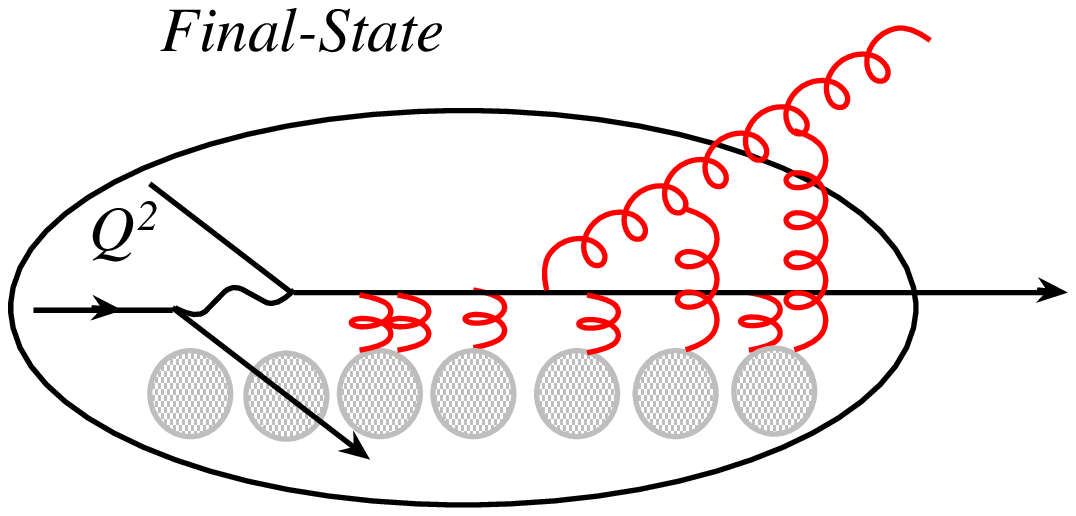,width=2.8in}
\psfig{file = 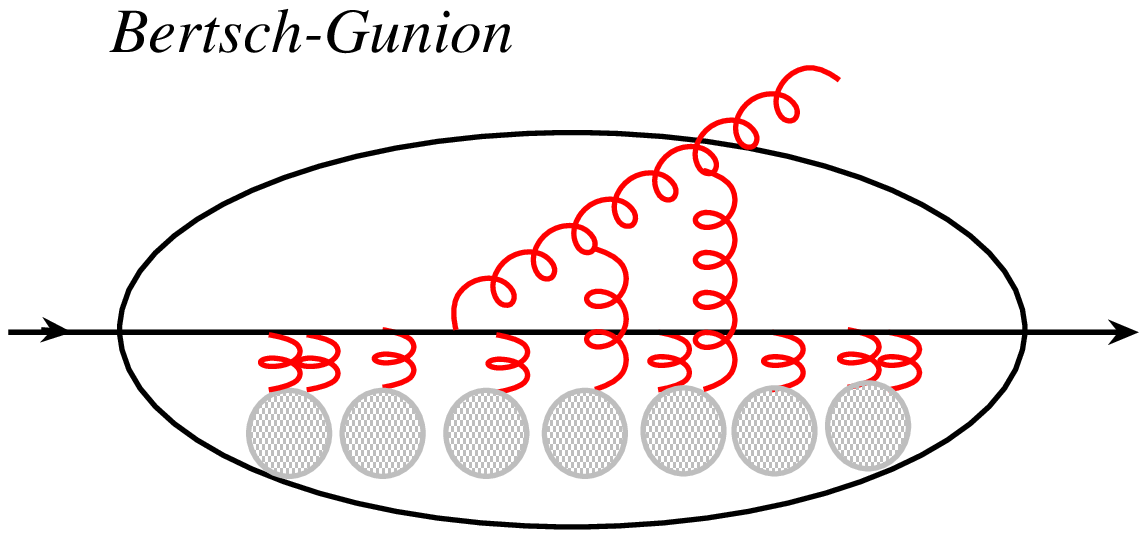,width=2.8in}
\caption{ Three distinct cases of medium-induced bremsstrahlung 
discussed in Ref.~\cite{Vitev:2007ve}  are illustrated. 
Top left: initial-state  energy loss
in the nucleus followed by a large $Q^2$ process,
resulting in the production of high-$p_T$ or high-$E_T$ 
particles and jets. Top right:  final-state  energy loss in the
nucleus, after a hard collision. 
Bottom: the case of asymptotic on-shell Bertsh-Gunion beam jets 
that do not undergo a hard scattering process is also shown.
}
\label{fig-schem}
\end{center} 
\end{figure}

In inclusive jet production in p+A collisions, the energetic partons before and after the hard collisions undergo multiple scattering and lose energy due to medium induced gluon bremsstrahlung. These processes are referred to as initial-state and final-state parton energy loss, respectively. A derivation to all orders in the correlations between the multiple scattering 
in nuclear matter of initial-state and final-state energy loss was performed in Ref.~\cite{Vitev:2007ve} using the Gyulassy-Levai-Vitev (GLV) reaction operator approach~\cite{Gyulassy:2000fs}, see Fig.~\ref{fig-schem} for illustration of those cases. The bremsstrahlung pattern of Bertsch-Gunion beam jets, partons that only undergo soft interactions in nuclear matter, was also obtained.   
It was demonstrated that at high energies the fractional energy loss $\Delta E/E$ of beam jets is the largest. The resulting  gluon bremsstrahlung may be partly or fully responsible for the
soft particle production in p+A and A+A reactions and the observed azimuthal asymmetries~\cite{Gyulassy:2014cfa,Biro:2015iua}, however it is not relevant  to high
$p_T$ particle and jet production.  For large $Q^2$ processes shown in Fig.~\ref{fig-schem}, initial-state energy loss is much larger and dominates over the final-state energy loss
when $E\rightarrow \infty$. Note that in phenomenological applications of cold nuclear matter energy loss  one must consider the large target nucleus  rest frame, where the energies of the incident and outgoing parton are very high~\cite{Vitev:2007ve} and, thus, final-state cold nuclear matter energy loss can indeed be neglected. Because of this, we will focus on the 
initial-state  energy loss as it is central to our paper.

The starting point of initial-state cold nuclear matter energy loss computation is the integral form of the  double differential medium-induced gluon bremsstrahlung spectrum~\cite{Vitev:2007ve} to first order in opacity:
\bea
k^+\frac{d^3N^g}{dk^+d^2k_{\perp}} = \frac{C_R\alpha_s}{\pi^2}\int d^2q_{\perp}\frac{\xi_{\rm eff}^2}{\pi (q_{\perp}^2+\xi^2)^2}
\left[\frac{L}{\lambda_g}\frac{q_{\perp}^2}{k_{\perp}^2(k_{\perp}-q_{\perp})^2}
-2\frac{q_{\perp}^2-q_{\perp}\cdot k_{\perp}}{k_{\perp}^2(k_{\perp}^2-q_{\perp}^2)}
\frac{k^+}{k_{\perp}^2\lambda_g}{\rm sin}\left(\frac{k_{\perp}^2L}{k^+}\right)\right],
\label{eq-Ng}
\eea
where $k^+$ is the large light cone momentum of the radiated gluon, $k_{\perp}$ is the transverse momentum relative to the direction of the energetic parent parton, and $q_{\perp}$ is the transverse momentum transfer between the propagating parton and the nuclear medium of size $L$ and uniform density. $C_R$ is the quadratic Casimir in the fundamental and adjoint representations of SU(3) for quarks and gluons, respectively. $\lambda_g$ is the gluon scattering length of ${\mathcal O}$(1 fm) and $\xi$ is the typical transverse momentum transfer per interaction representing the interaction strength between the propagating parton and medium. On the other hand,  the finite range of integration $q_\perp^2 \leq Q^2_{\rm max} = \xi E_{\rm jet}/2$ leads to $\xi_{\rm eff}^2 = \xi^2 (\xi^2 + Q_{\rm max}^2)/Q_{\rm max}^2$~\cite{Vitev:2007ve}. Effective parton mass due to interactions in matter is 
included as $k_{\perp}^2 \rightarrow k_{\perp}^2 + \xi^2$.  
The gluon number/intensity spectra  and  average gluon number/parton energy loss are then obtained as ($k^+ \approx 2 \omega$):
\bea
\label{intensity}
& \frac{dN^g}{d\omega}  \approx 2 \frac{dN^g}{d k^+} = 2 \int d^2 k_\perp  \,  \left.\frac{d^3N^g}{dk^+d^2k_{\perp}}\right|_{>0} \, , \qquad   
  \omega \frac{dN^g}{d\omega}  \approx k^+ \frac{dN^g}{d k^+} = \int d^2 k_\perp  \,  \left.k^+\frac{d^3N^g}{dk^+d^2k_{\perp}}\right|_{>0}  \, , 
\\
& N^g  =  \int d \omega  \,  \frac{dN^g}{d\omega}   \, , \qquad       
   \Delta E  = \int d\omega   \,   \omega \frac{dN^g}{d\omega}      \,  .  
\label{eloss}
\eea

In heavy ion collisions, where final-state energy loss effects dominate, phenomenological applications have improved upon the use of the mean energy loss
$\Delta E$ in Eq.~(\ref{eloss}).  This approach can be generalized to initial-state energy loss. With the differential distribution of  radiated gluons $dN^g/d\omega$  in Eq.~\eqref{intensity} at hand, one can calculate the probability density $P_{q,g}(\epsilon)$ for quarks and gluons to lose a fraction of their energy $\epsilon=\sum_i \Delta E_i/E$ due to multiple gluon emission in the Poisson approximation. 
 We can further evaluate the mean energy loss fraction as follows:
\bea
\langle\epsilon_{q,g} \rangle=\left\langle\frac{\Delta E_{q,g~ \rm initial-state}}{E} \right\rangle = \int_0^1 d\epsilon \,\epsilon\, P_{q,g}(\epsilon), \quad { \rm where} \quad 
\int_0^1d\epsilon \, P_{q,g}(\epsilon)=1 \; .
\label{eq:fraction}
\eea
Note that in Eq.~(\ref{eq:fraction})  with the subscripts $q,\,g$ we explicitly indicate that quarks and gluons radiate different numbers of gluons and lose different fractions of their energy. This was implicit in Eqs.~(\ref{intensity}), (\ref{eloss}).
If the incident parton $a$ (either quark or gluon) loses a fractional energy $\epsilon$, to satisfy the same final-state kinematics, it must carry a larger fraction of the colliding hadron momentum and, in turn, a larger value of momentum fraction $x_a$. This can be implemented in the cross section calculations in Eq.~\eqref{eq:p+p} in the following way~\cite{Kang:2012kc}:
\bea
& f_{q/p}(x_a, \mu) \rightarrow \int_0^1 d\epsilon \, P_{q}(\epsilon)\, f_{q/p}\left(\frac{x_a}{1-\epsilon},\mu\right), \quad
f_{g/p}(x_a, \mu) \rightarrow \int_0^1 d\epsilon \, P_{g}(\epsilon)\, f_{g/p}\left(\frac{x_a}{1-\epsilon},\mu\right).
\eea
Such an implementation has taken into account the fluctuation in the cold nuclear matter energy loss via $P_{q,g}(\epsilon)$, and can be computationally demanding in the calculations of the nuclear modification factors in the next section. The result of energy loss fluctuations is a reduction in  the attenuation of the particle and jet cross sections, effectively corresponding to a smaller fractional energy loss than that in Eq.~(\ref{eq:fraction}).   To speed up the evaluation, we implement this effect as a momentum fraction shift in the PDFs,
\bea
f_{q/p}(x_a,\mu) \rightarrow f_{q/p}\left(\frac{x_a}{1-\epsilon_{q,\rm eff}},\mu\right),
\qquad
f_{g/p}(x_a,\mu) \rightarrow f_{g/p}\left(\frac{x_a}{1-\epsilon_{g,\rm eff}},\mu\right),
\label{eq-mPDFs}
\eea
where we follow Refs.~\cite{Sharma:2009hn,Kang:2012kc} and use $\epsilon_{q,g,\rm eff} = 0.7\,\langle\epsilon_{q,g}\rangle$ with the mean energy loss fraction $\langle\epsilon_{q,g}\rangle$ given by Eq.~\eqref{eq:fraction}.  This result immediately implies that the nuclear modification of single inclusive jet production in p+A collisions depends not only on the magnitude of initial-state cold nuclear matter energy loss, but also on the steepness of the parton distribution functions. In particular, large suppression is expected for  jet production at forward rapidity and large transverse momentum, where the parton momentum fraction $x_a$ in the proton is large and $f_{q,g/p}(x_a,\mu)$ is steeply falling. We will see these effects explicitly in the next section, where we evaluate the dependence of cold nuclear matter energy loss effects on the jet rapidity $y$ and transverse momentum $p_T$.
\begin{figure}[b!]
\psfig{file=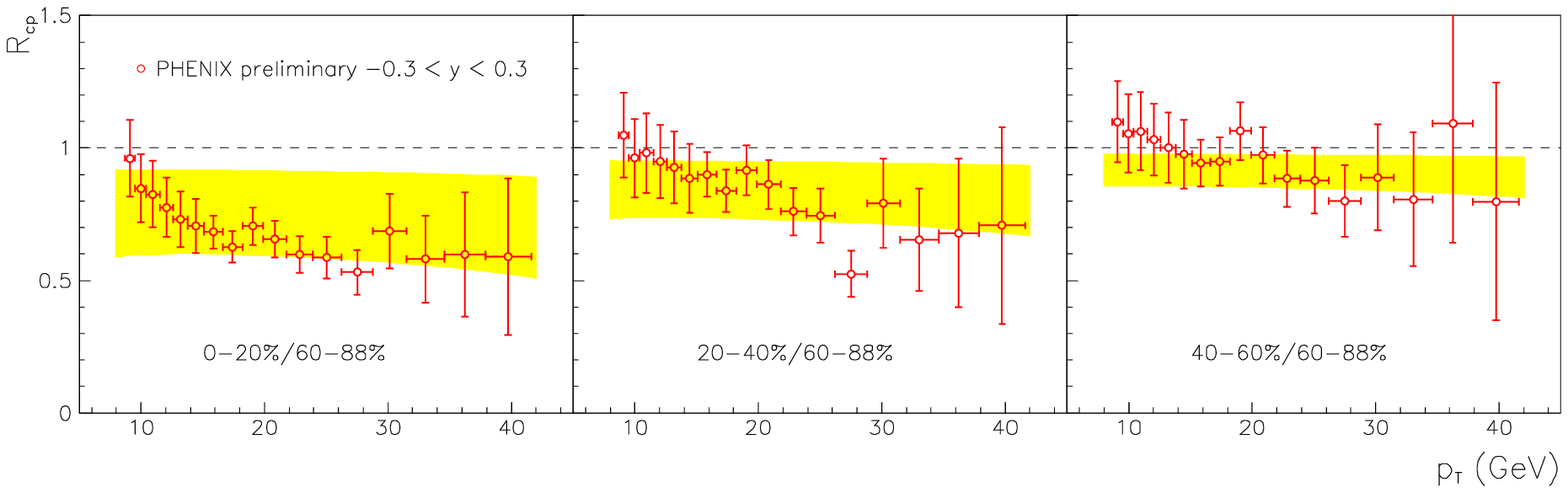, width=5.8in}
\caption{$R_{\rm cp}$ for inclusive jet production in d+Au collisions at $\sqrt{s}=200$ GeV in central (left), mid-central (middle) and mid-peripheral (right) events. Data are from PHENIX \cite{Perepelitsa:2013jua}. Yellow bands correspond to the theoretical uncertainty of initial-state parton energy loss in the range $0.175 < \xi< 0.7$ GeV.}
\label{fig-Rcp_dAu}
\eef
\begin{figure}[b!]
\psfig{file=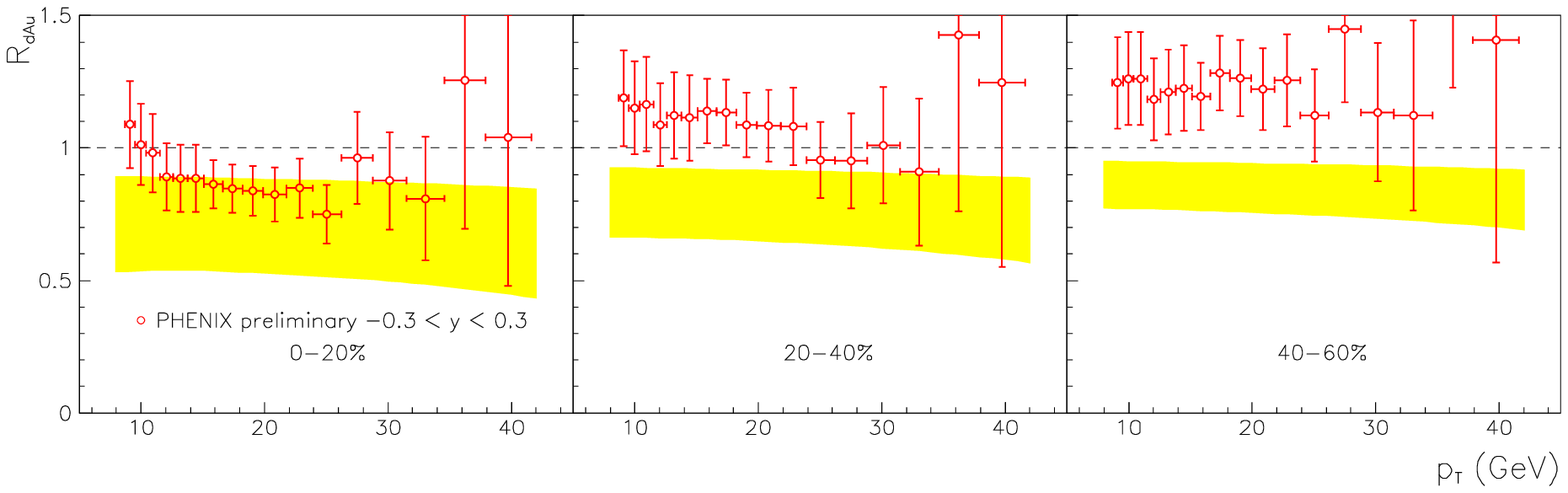, width=5.8in}
\caption{Same as in Fig.~\ref{fig-Rcp_dAu}, but for $R_{\rm dAu}$.}
\label{fig-dAu}
\eef

\section{Numerical results at RHIC and LHC}
\bef
\psfig{file=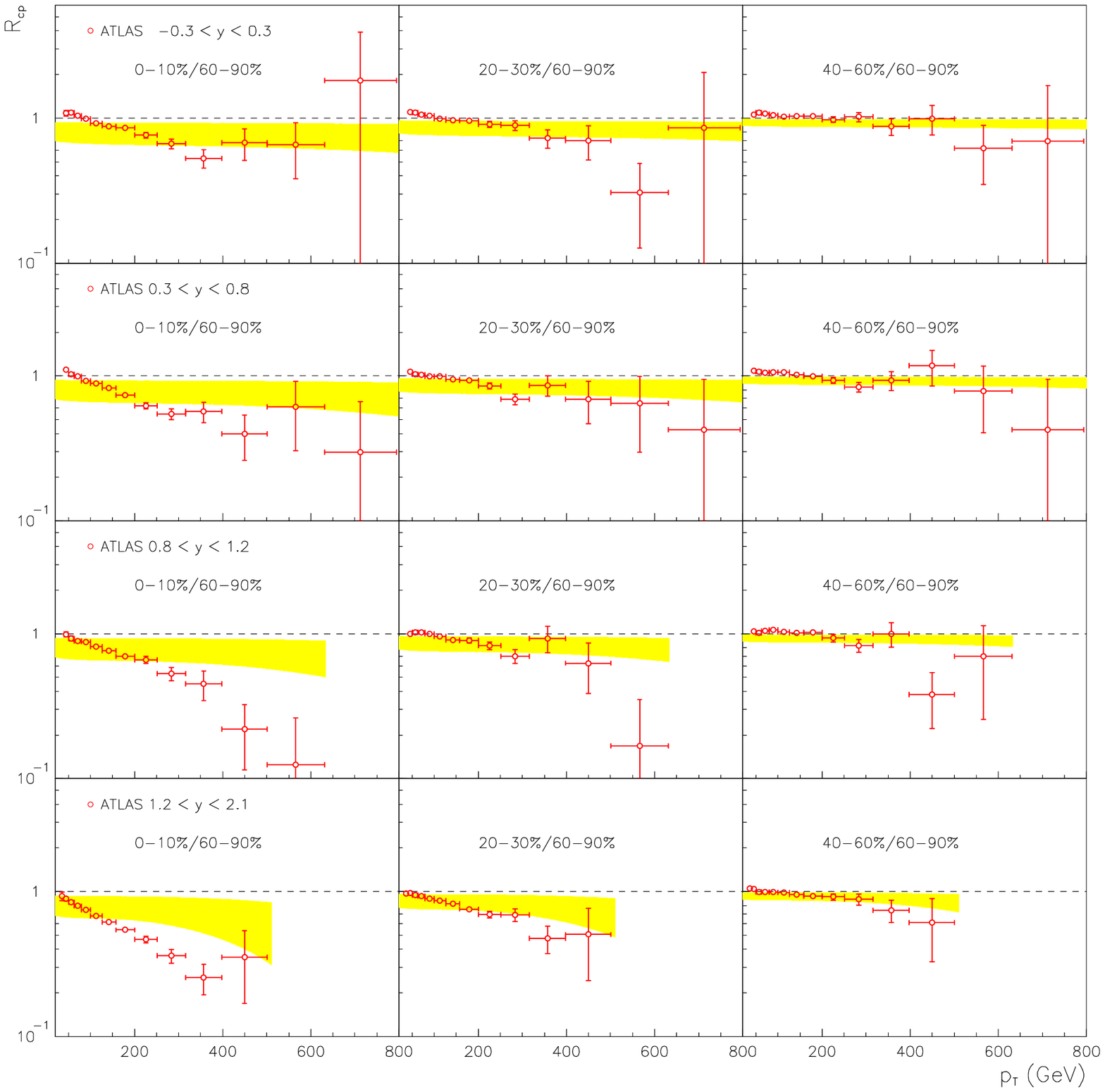, width=5.8in}
\caption{$R_{\rm cp}$ for inclusive jet production in p+Pb collisions at $\sqrt{s}=5.02$ TeV in central (left), mid-central (middle) and mid-peripheral (right) events. Four different rapidity intervals ($-0.3<y<0.3$, $0.3<y<0.8$, $0.8<y<1.2$ and $1.2<y<2.1$) are presented. Data are from ATLAS collaboration at the LHC~\cite{ATLAS:2014cpa}.}
\label{fig-Rcp_pt1}
\eef

In this section we present numerical calculations of initial-state parton energy loss effects on inclusive jet production in p+A collisions  and compare to experimental data from d+Au collisions at RHIC and p+Pb collisions at the LHC. To start, we define the usual nuclear modification factor $R_{\rm pA}$:
\bea
R_{\rm pA}=\frac{1}{\langle N_{\rm coll} \rangle} \frac{d\sigma^{pA}_{\rm jet}/dyd^2p_T}{d\sigma^{pp}_{\rm jet}/dyd^2p_T},
\label{eq-RpA}
\eea
where $\langle N_{\rm coll} \rangle$  is the average number of binary nucleon-nucleon collisions and is computed at a given centrality, and $d\sigma^{pA}_{\rm jet}/dyd^2p_T$ and $d\sigma^{pp}_{\rm jet}/dyd^2p_T$ represent the differential cross section of jet production in p+A and p+p collisions, respectively. The deviation of $R_{\rm pA}$ from unity reveals non-trivial nuclear effects in p+A collisions. Following experimental measurements, cold nuclear matter effects can also be quantified by the central-to-peripheral ratio $R_{\rm cp}$:
\bea
R_{\rm cp}=\frac{1}{R_{\rm coll}}\frac{d\sigma^{pA}_{\rm jet}/dyd^2p_T\big|_{\rm cent}}{d\sigma^{pp}_{\rm jet}/dyd^2p_T\big|_{\rm per}},
\label{eq-Rcp}
\eea
where $R_{\rm coll}$ represents the ratio of $N_{\rm coll}$ in a given centrality interval to that in the most peripheral interval.  It is given by the expression $R_{\rm coll}=\langle N_{\rm coll}^{\rm cent}\rangle/\langle N_{\rm coll}^{\rm per}\rangle$.

In both p+p and p+A collisions, we use CTEQ6L1 parton distribution functions~\cite{Pumplin:2002vw}. We choose the factorization and renormalization scales to be equal in the rest of this section and set them as $\mu=p_T$. In the numerical evaluation of initial-state parton energy loss, we use $\lambda_g=1$~fm for the gluon mean free path and vary the interaction strength between the propagating jet and the QCD medium by changing the typical momentum transfers $\xi$ in the range of 0.175 to 0.7 GeV. Such a range has been shown to give a good description for the nuclear modification of Drell-Yan dilepton production in fixed target experiments \cite{Neufeld:2010dz}. 
\begin{figure}[t!]
\psfig{file=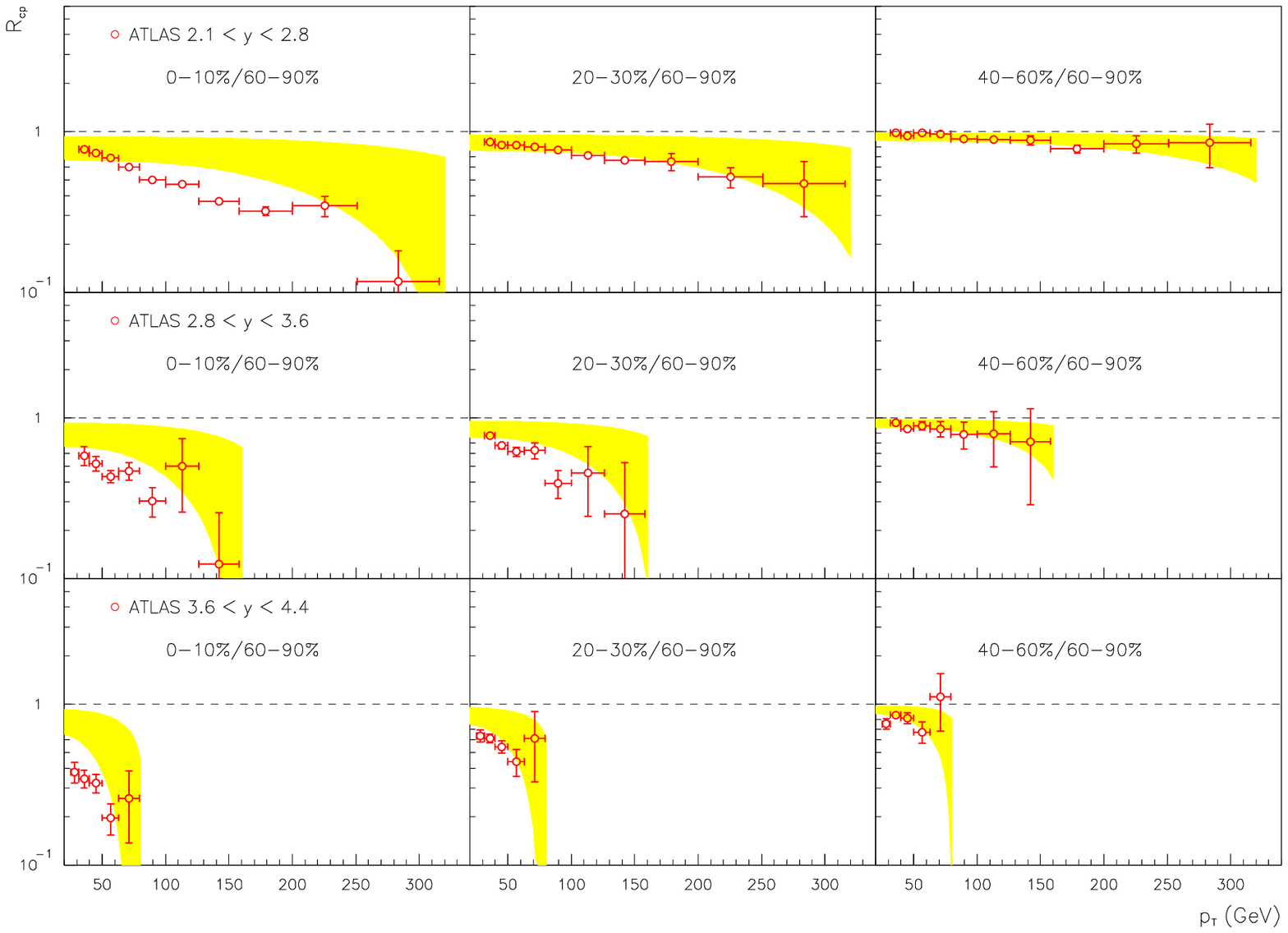, width=5.8in}
\caption{Same as Fig.~\ref{fig-Rcp_pt1}, but for the three other rapidity intervals ($2.1<y<2.8$, $2.8<y<3.6$ and $3.6<y<4.4$).}
\label{fig-Rcp_pt2}
\eef
\begin{figure}[t!]
\psfig{file=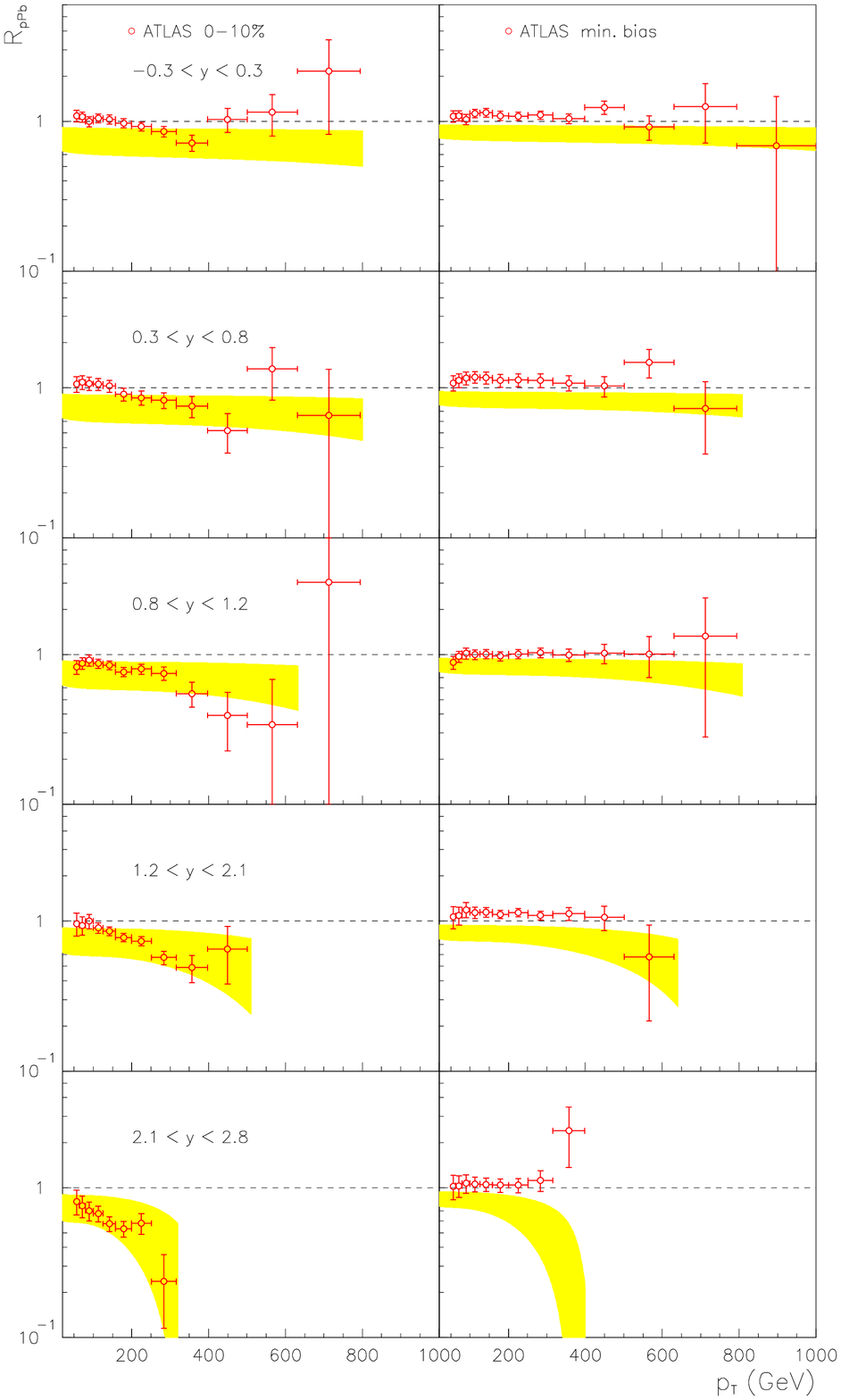, width=4.1in}
\caption{$R_{\rm pPb}$ for inclusive jet production in p+Pb collisions at $\sqrt{s}=5.02$ TeV in central (left) and minimum bias (right) events. Five different rapidity intervals ($-0.3<y<0.3$, $0.3<y<0.8$, $0.8<y<1.2$, $1.2<y<2.1$ and $2.1<y<2.8$) are presented. Data is from ATLAS collaboration at the LHC~\cite{ATLAS:2014cpa}.}
\label{fig-RpPb}
\eef
\begin{table}[h!]
\begin{center}
\begin{tabular}{|c|c|c|c|c|c|}
\hline
\multirow{3}{*}{d+Au} 
& \;  Centrality   \;   &  $0 - 20\% $ &   $20-40 \% $  &  $40 - 60\% $ & \;  $60-88\%$  \;  \\
\cline{2-6}
& $\langle N_{\rm coll}\rangle$  &  15.1  &  10.2  &  6.6  & 3.2  \\
\cline{2-6}
&  $L$ [fm]  &  10.2 & 6.8 & 4.1  & 1.5 \\
\hline
\hline
\multirow{3}{*}{p+Pb} 
& \;  Centrality   \;   &  $0 - 10\% $ &   $20-30 \% $  &  $40 - 60\% $ & \;  $60-90\%$  \;  \\
\cline{2-6}
& $\langle N_{\rm coll}\rangle$  &  14.6  &  10.4  &  6.4  & 3.0  \\
\cline{2-6}
& $L$ [fm]  &  10.6 & 7.3 & 4.2  & 1.6 \\
\hline
\end{tabular}
\end{center}
\caption{\label{table-L} Estimated effective path length $L$ for different centralities in d+Au (top table) collisions at RHIC $\sqrt{s} = 200$ GeV and p+Pb (bottom table) collisions at the LHC $\sqrt{s} = 5.02$ TeV. $\langle N_{\rm coll}\rangle$ for RHIC and LHC are consistent with those from PHENIX collaboration at \cite{Adare:2013nff} and ATLAS collaboration at \cite{ATLAS:2014cpa}, respectively.} 
\end{table}

To calculate the centrality dependence of cold nuclear matter energy loss in p+A collisions, we use a Glauber model~\cite{Miller:2007ri,d'Enterria:2003qs} to determine the effective path length $L$ in Eq.~\eqref{eq-Ng} for jet production at a given centrality class, in which we choose the inelastic nucleon-nucleon scattering cross section $\sigma_{in} = 42$ mb at RHIC energy of $\sqrt{s} = 200$ GeV and $\sigma_{in} = 70$ mb at the LHC energy of $\sqrt{s} = 5.02$ TeV. The calculated effective path length $L$ for the corresponding average number of binary collisions $\langle N_{\rm coll} \rangle$ are summarized in Table.~\ref{table-L}, for both d+Au collisions at RHIC and p+Pb collisions at the LHC. $\langle N_{\rm coll}\rangle$ for RHIC and LHC are taken from the PHENIX collaboration~\cite{Adare:2013nff} and the ATLAS collaboration~\cite{ATLAS:2014cpa}, respectively.

We are now ready to compare our theoretical results with experimental data. In Fig.~\ref{fig-Rcp_dAu} we plot the central-to-peripheral ratio $R_{\rm cp}$ for jet production in d+Au collisions at $\sqrt{s}=200$ GeV as a function of the jet transverse momentum $p_T$, where the peripheral class is taken to be $60-88\%$ centrality. The experimental data are from the PHENIX collaboration~\cite{Perepelitsa:2013jua} at central rapidity $-0.3<y<0.3$ in three different centrality intervals: $0-20\%$ (left), $20-40\%$ (middle), $40-60\%$ (right), respectively~\footnote{Finalized d+Au data appeared after our work was completed. Comparison to our theoretical calculations can be found in the PHENIX paper~\cite{Adare:2015gla}.}. The yellow bands are our theoretical calculations, representing the uncertainty of typical momentum transfers between the propagating parton and nuclear medium in the range $0.175 < \xi < 0.7$ GeV as mentioned above. This momentum range  has to be further constrained by unambiguous experimental measurements, e.g., the anticipated E906 Drell-Yan results from Fermilab fixed target experiments. As we have argued, cold nuclear matter energy loss leads to suppression of the jet production cross section. This attenuation becomes stronger in the most central collisions due to the increase in the effective path length $L$, as shown in Eq.~\eqref{eq-Ng}. Our numerical calculations give a reasonable description of the PHENIX data for $R_{\rm cp}$. However, as shown in Fig.~\ref{fig-dAu}, it fails to describe the nuclear modification factor $R_{\rm dAu}$ for non-central collisions. In particular the enhancement in the peripheral collisions at $40-80\%$ centrality class is not described in this picture. 

With the same set of gluon scattering length $\lambda_g$ and in-medium momentum transfers $\xi$ in cold nuclear matter, we summarize our theoretical calculations of $R_{\rm cp}$ in p+Pb collisions at LHC $\sqrt{s}=5.02$ TeV as a function of $p_T$ in Figs.~\ref{fig-Rcp_pt1} and \ref{fig-Rcp_pt2}, where the peripheral class is taken to be $60-90\%$ centrality. The experimental data are from ATLAS collaboration at the LHC~\cite{ATLAS:2014cpa}. Seven rapidity intervals, $-0.3<y<0.3$, $0.3<y<0.8$, $0.8<y<1.2$, $1.2<y<2.1$, $2.1<y<2.8$, $2.8<y<3.6$ and $3.6<y<4.4$ in three different centrality bins ($0-10\%$, $20-30\%$, $40-60\%$) are considered. Our theoretical calculations give a reasonable description for most of the data, and provide support for the cold nuclear matter energy loss picture. In particular, ATLAS data show a clear dependence of nuclear modification on the centrality, which seem to follow our expectation based on effective path length dependence in the cold nuclear matter energy loss formalism~\cite{Vitev:2007ve,Neufeld:2010dz}. In all centrality intervals the central-to-peripheral ratio $R_{\rm cp}$ decreases with increasing transverse momentum $p_T$ and rapidity $y$. Such dependences also find natural explanation in the cold nuclear matter energy loss picture. As we have emphasized in the last section, energy loss is very sensitive to the steepness of PDFs of the projectile proton. As $p_T$ and $y$ increase, the sampled parton momentum fraction in the projectile proton $x_a$ becomes larger, so does the the  slope of the PDFs  $|df_{a/p}(x_a, \mu)/dx_a|$, and, thus, a larger suppression is expected. We notice here that in the highest $p_T$ bins of the most central collisions in the most forward rapidity region, the nuclear suppression $R_{\rm cp}$ can be as low as $0.2$. This implies that initial-state cold nuclear matter effects in inclusive jet production may also be  significant in A+A reactions in this kinematic region~\cite{Vitev:2009rd}.

We further plot the nuclear modification factor $R_{\rm pPb}$ in the most central ($0-10\%$) (left) as well as minimum bias (right) collisions in Fig.~\ref{fig-RpPb}. The same patterns as in $R_{\rm cp}$ and reasonable agreements between our theoretical calculations and experimental data are found. The agreement is best in central collisions, shown in the left panels
of  Fig.~\ref{fig-RpPb}. A nuclear PDF-based calculation does not reproduce this suppression.  However, similar to  RHIC findings, anomalous nuclear enhancement of single jet production in peripheral collisions is also observed at the LHC. Such an enhancement is again beyond what we can expect from either the initial-state parton energy loss picture or the nPDF model. In minimum bias collisions, shown in the right panels of  Fig.~\ref{fig-RpPb}, there is no clear indication that $R_{\rm pPb}$ deviates form unity, as expected in energy loss calculations. This is consistent with the nPDF large $Q^2$ behavior. Within statistical and systematic uncertainties, the data is, however, also compatible with the upper edge of the CNM  energy loss band.

\bef
\psfig{file=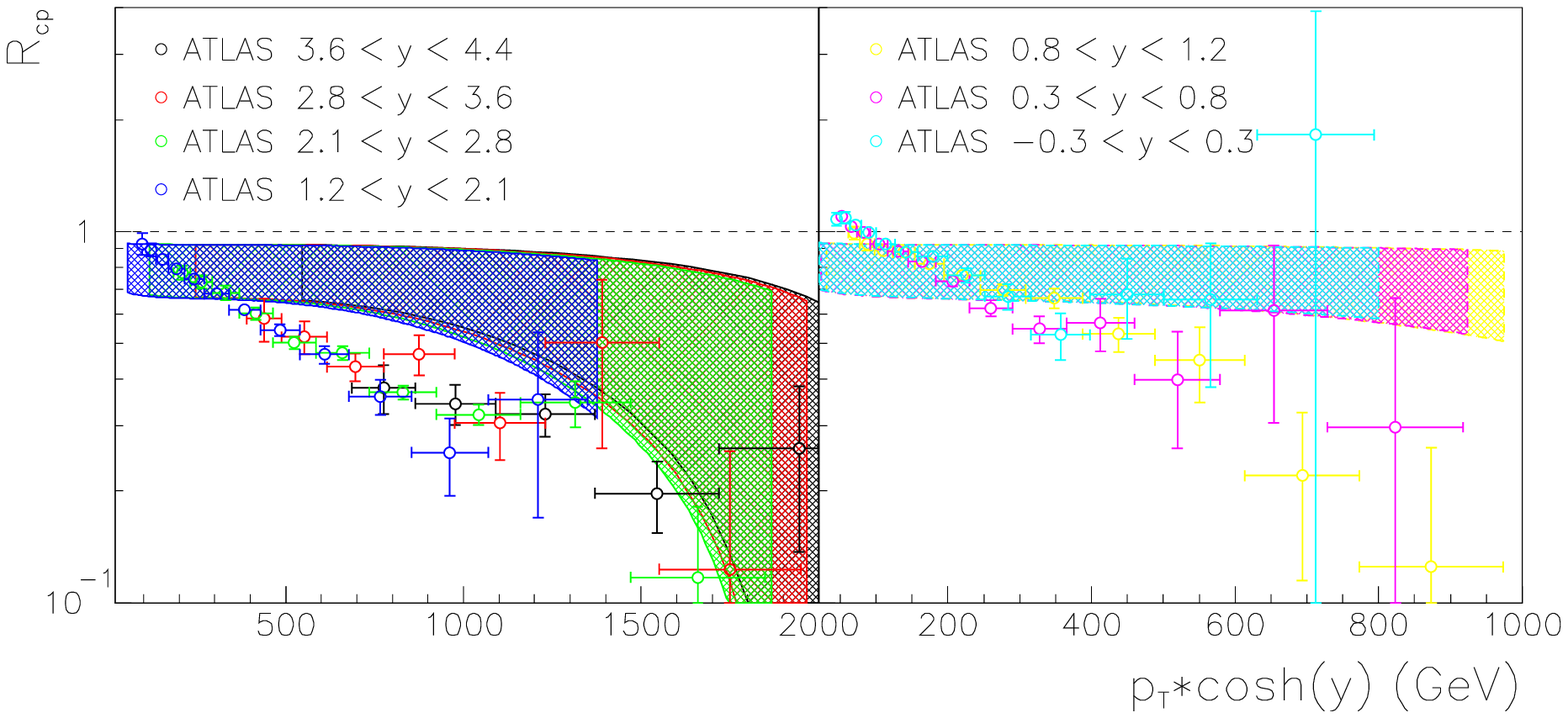, width=5.1in} \\ \vspace*{0.2in}
\psfig{file=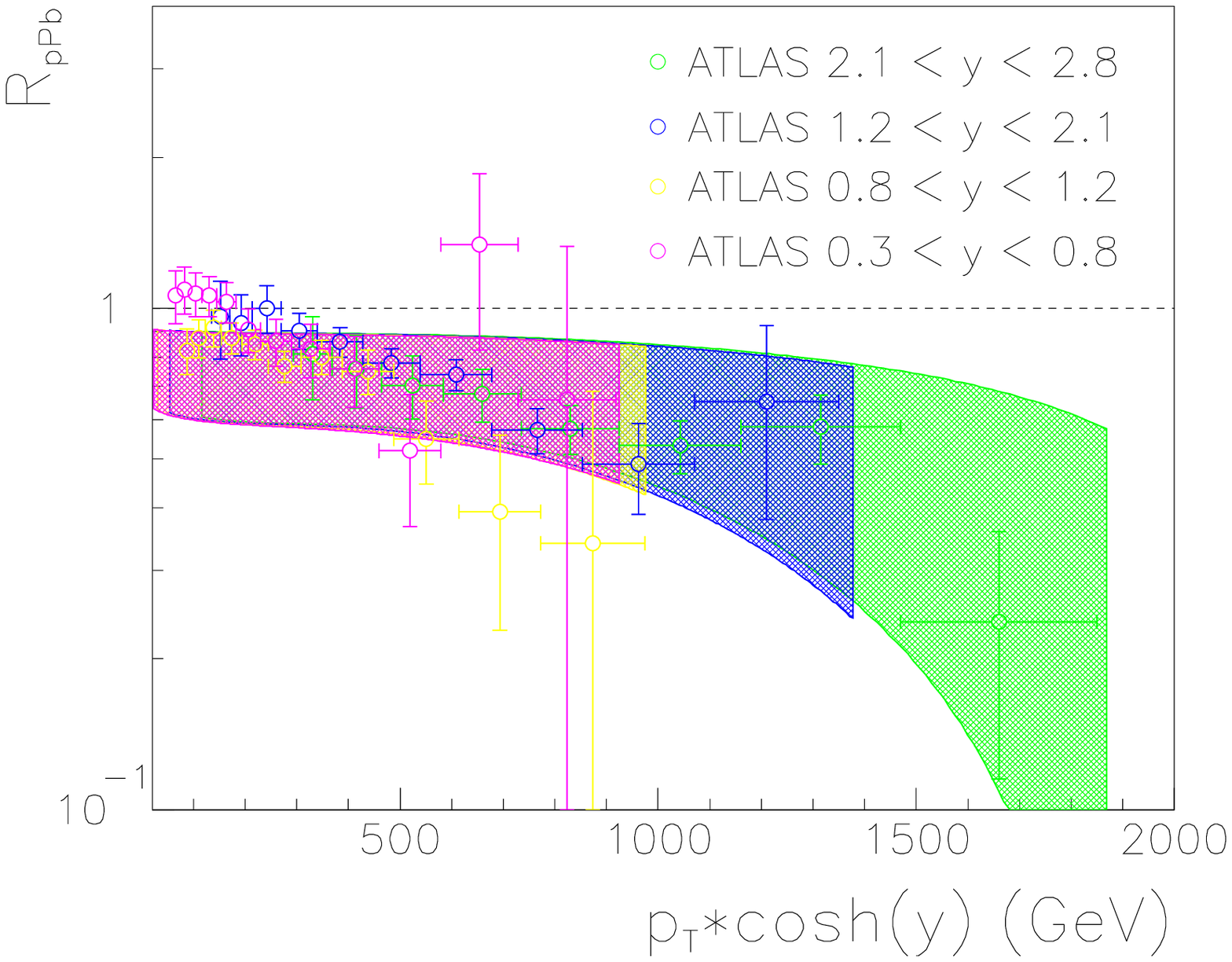, width=3.2in}
\caption{$R_{\rm cp}$ (top) and $R_{\rm pPb}$ (bottom) in $0-10\%$ p+Pb collisions as a function of $p_T\times \cosh (y)$  are shown at different rapidities in the most central collisions ($0-10\%$). Seven rapidity intervals are presented for $R_{\rm cp}$ and four for $R_{\rm pPb}$. Data is from ATLAS collaboration at the LHC~\cite{ATLAS:2014cpa}.}
\label{fig-pty}
\eef

In search of a universal relationship between the nuclear modification factors in different rapidity ranges ATLAS has further plotted $R_{\rm cp}$ and $R_{\rm pPb}$ as a function of the jet kinematic variable $p_T\times \cosh(y)$~\cite{ATLAS:2014cpa}, which is approximately the total energy of the jet. They find that, when plotted against this variable, the nuclear modification factors in the {\it central} and {\it forward} rapidity regions fall along the same curve. However, $R_{\rm cp}$ and $R_{\rm pPb}$ at {\it backward} rapidity do not follow this trend. These patterns suggest that the observed nuclear modification may depend on the initial parton kinematics, especially the parton momentum fraction $x_a$ in the projectile proton. Such a behavior is exactly what one can expect in the initial-state parton energy loss picture, which we will now elaborate in more detail. From the  partonic kinematics at LO  one can easily derive in the forward rapidity region, where  $x_a \gg x_b$,  
\bea
p_T\times \cosh(y) \propto x_a\,\sqrt{s}  \, ,
\label{eq:xab}
\eea
and we have used the fact that the cross section is dominated by configurations where the observed and recoil jets are at approximately equal rapidities. 
 Therefore, the kinematic variable $p_T\times \cosh(y)$ introduced by ATLAS  is simply the projectile parton momentum fraction $x_a$ times a constant factor.  Since initial-state parton energy loss is equivalent to a shift in $x_a$, as shown in Eq.~\eqref{eq-mPDFs}, the nuclear modification factors will follow the same trend once they are plotted as a function of $x_a$. This is precisely what we have observed in Fig.~\ref{fig-pty}, where we show the comparison between our theoretical calculations of $R_{\rm cp}$ and $R_{\rm pPb}$ and the experimental data, against the quantity $p_T\times \cosh(y)$. As one can see, all theoretical calculations for different rapidity intervals are on top of each other, following the correct scaling with $x_a$. There are clearly some quantitative difference between our theoretical calculations and experimental data, especially in $R_{\rm cp}$, which we attribute partly to the unexplained anomalous enhancement of jet production in peripheral collisions.  With this caveat, the reasonable agreement on both the trend and the size of the nuclear modification in p+A collisions is still very encouraging. On the other hand, in the backward rapidity region $x_b$  is comparable to (even larger than) $x_a$, and one does not have a simple 
proportionality between $x_a$ and the total jet energy.   Thus, a scaling behavior, as observed in forward rapidity region, should not be expected. This is exactly what ATLAS collaboration has observed.

\section{Summary}
In this paper we studied to what extent the nuclear modification of inclusive jet production in p+A collisions can be described by  standard cold nuclear matter effects. In particular, for jet production at large transverse momentum $p_T$ we evaluated the nuclear modification factor $R_{\rm pA}$ and the central-to-peripheral ratios $R_{\rm cp}$ of inclusive jet production due to  initial-state cold nuclear matter energy loss in d+Au collisions at RHIC and p+Pb collisions at the LHC. We found that our theoretical calculations can capture quantitatively the bulk of the observed modifications for both RHIC and LHC experiments from central to semi-central collisions. The agreement is good in central reactions and   the upper edge of the CNM energy loss band is still consistent with data  in minimum bias reactions if the statistical and systematic uncertainties are taken into account. On the other hand, a nPDF-based calculation is consistent with the minimum bias data in inclusive jet production form RHIC and LHC, but fails to describe central collisions.  What is particularly important to note is that the observed scaling patterns of $R_{\rm cp}$ and $R_{\rm pPb}$ against the total energy of the jet $p_T\times \cosh(y) \propto x_a$ (the parton momentum fraction in the projectile proton) in the forward rapidity region in p+Pb collisions find their natural explanation in the picture of cold nuclear matter energy loss. We further note that $x_a \approx x_F$, Feynman $x$, at forward rapidity and similar scaling of the nuclear attenuation is expected for large $x_F$. It has indeed been observed for different final states~\cite{Leitch:1999ea}.    
The encouraging comparison between theoretical calculations and experimental data indicates the significance of cold nuclear matter energy loss for understanding particle and jet production in p+A collisions.

On the other hand, the observed nuclear enhancement in peripheral collisions is difficult to understand in our picture or the nPDF model. Such an enhancement might have a different origin, for example  from ``centrality bias'', and needs to be explored further. So far it has no first-principles explanation at the quantitative level. The anomalous enhancement of jet production also affects the central-to-peripheral  ratio and is partly responsible for the small values of  $R_{\rm cp}$.
 Experimentally, future results from RHIC, LHC and the fixed target p+A program at Fermilab will be essential to further elucidate the non-trivial nuclear modification of high $p_T$/high invariant mass particles and jets in p+A reactions  and  to constrain the  transport properties of cold nuclear matter  in the current energy loss formalisms.
Theoretically, it will be important to understand if there is  centrality selection bias and what is its dynamical origin. 
As far as inelastic initial-state interactions are concerned, a step forward will  be to go beyond the soft-gluon energy
loss approximation and obtain the full medium-induced splitting kernels~\cite{Ovanesyan:2011kn}. With such cold nuclear matter splitting kernels at hand, one can treat vacuum and in-medium parton showers 
on the same footing, following the progress recently made on the implementation of final-state QGP effects~\cite{Kang:2014xsa,Chien:2015vja,Chien:2015hda}. We leave such developments  for future work.

\section*{Acknowledgments}
We thank D.V. Perepelitsa for providing us data of jet production in d+Au and p+Pb collisions measured by PHENIX and ATLAS collaboration, respectively. This work is supported by the US Department of Energy, Office of Science under  Contract No. DE-AC52-06NA25396, by the DOE Early Career Program, and in part by the LDRD program at Los Alamos National Laboratory.

\bibliographystyle{h-physrev5}   
\bibliography{biblio}

\end{document}